\pdfoutput=1
\documentclass[aps,%
 reprint,
superscriptaddress,
 amsmath,amssymb,
]{revtex4-1}

\usepackage{graphicx}
\usepackage{dcolumn}
\usepackage{bm}
\usepackage{mathrsfs} 
\usepackage{stmaryrd} 
\usepackage{multirow} 
\usepackage{hyperref}
\usepackage{color} 

\newcommand{\aap}{{Astron. Astrophys. }}

\newcommand{\aj}{{Astron. J. }}
\newcommand{\apss}{{Astrophys. and Space Science }}
\newcommand{\up}[1]{\textsuperscript{#1}}

\begin{document}

\preprint{APS/123-PRD}

\title{Regression analysis with missing data and unknown colored noise: application to the MICROSCOPE space mission}

\author{Quentin Baghi}
\email{quentin.baghi@onera.fr}
\affiliation{
ONERA - The French Aerospace Lab,
29 avenue de la Division Leclerc, 92320 Chatillon, France \\
}
\author{Gilles M\'{e}tris}%
\email{gilles.metris@oca.eu}
\affiliation{
 Geoazur (UMR 7329), Observatoire de la C\"{o}te d'Azur
 Bât 4, 250 rue Albert Einstein, Les Lucioles 1, Sophia Antipolis, 06560 Valbonne, France \\
}
\author{Jo\"{e}l Berg\'{e}}
\author{Bruno Christophe}
\author{Pierre Touboul}
\author{Manuel Rodrigues}
\affiliation{
ONERA - The French Aerospace Lab,
29 avenue de la Division Leclerc, 92320 Chatillon, France \\
}

\date{February 17, 2015}

\begin{abstract}
The analysis of physical measurements often copes with highly correlated noises and interruptions caused by outliers, saturation events or transmission losses. 
We assess the impact of missing data on the performance of linear regression analysis involving the fit of modeled or measured time series. We show that data gaps can significantly alter the precision of the regression parameter estimation in the presence of colored noise, due to the frequency leakage of the noise power. We present a regression method which cancels this effect and 
estimates the parameters of interest with a precision comparable to the complete data case, even if the noise power spectral density (PSD) is not known \textit{a priori}. The method is based on an autoregressive (AR) fit of the noise, which allows us to build an approximate generalized least squares estimator approaching the minimal variance bound. The method, which can be applied to any similar data processing, is tested on simulated measurements of the MICROSCOPE space mission, whose goal is to test the Weak Equivalence Principle (WEP) with a precision of $10^{-15}$. In this particular context the signal of interest is the WEP violation signal expected to be found around a well defined frequency.
We test our method with different gap patterns and noise of known PSD and find that the results agree with the mission requirements, decreasing the uncertainty by a factor 60 with respect to ordinary least squares methods. We show that it also provides a test of significance to assess the uncertainty of the measurement.

\begin{description}
\item[PACS numbers]
\item[Keywords]
{Data processing, Experimental test of gravitational theories, Spaceborne instruments}
\end{description}
\end{abstract}

\maketitle

\section{Introduction}
\label{intro}

Situations where series of measurements, ideally regularly sampled, suffer from short interruptions are common in a wide range of applications and experimental set-ups. It is also usual to perform linear regression analysis of data samples, in order to estimate parameters of interest by fitting other data series to the measured signals. In particular, this is a typical scenario for space missions in fundamental physics such as MICROSCOPE \cite{Touboul,Berge} and LISA Pathfinder \cite{PhysRevD.90.042003}. Long time integrations are needed by these experiments to reach the required signal-to-noise ratios (SNR) or the required levels of free-fall at the frequencies of interest. The duration of such measurements increases the probability to have invalid data in the integration period. It has been found that gaps could arise in the time series measured by the accelerometers carried on-board the MICROSCOPE satellite, and that those gaps could have substantial impact on the outcome of the regression when data is noisy. 

Here ``gaps'' refers to either lack of data or unusable information such as saturations and outliers during short or long time spans, which are eventually discarded. In the case of the MICROSCOPE space mission, discontinuities in the data availability could be due to data losses in the telemetry transmission, while data alteration could be the consequence of three main identified causes: crackles in the cold gas tanks triggered by decreasing pressure as they empty, crackles in the multi-layer insulation (MLI) coating due to temperature variations in flight, or micro-meteorites impacts. All saturated data are clearly identified by a flagging system in the telemetry. 

The objective of the MICROSCOPE signal processing can be regarded as rather general. It consists in detecting and estimating the amplitude of a periodic signal present in some measured time series. In the studied case the signal is the signature of a possible violation of the Weak Equivalence Principle (WEP), as detailed later, and is expected to arise around a certain frequency that we denote $f_{\rm EP}$. The amplitude to be estimated is the ``EP parameter'', denoted  $\delta$. In previous works \cite{hardy2013determination} the data analysis had been optimized in order to minimize the projection of possible unknown harmonic perturbations onto the signal of interest by an appropriate tuning of its frequency $f_{\rm EP}$ and/or the integration duration, in particular in the case of missing data. At the time, instrumental noise had been disregarded in order to exclusively deal with projection effects. Here we rather focus on the impact of missing data on the noise affecting the estimation. 

While the proposed approach is applied to MICROSCOPE simulated data, it leads to provide a robust method to estimate one or several deterministic components in the general context of time series with missing data affected by unknown colored noise. Although we have physical models of the expected noise spectrum, we assume in this study that it is not known \textit{a priori}, allowing us to cope with the most general situation. 

We show that noise distortions due to missing data points may dramatically increase the uncertainty of the estimation. This is due to the convolution effect between the observation window and the original noise spectrum, which leads to a leakage of the frequencies where the power is high to the frequencies where the power is low.

Methods such as Ordinary Least Squares (OLS) or equivalently Lomb-Scargle periodogram \cite{Lomb,Scargle}, as well as CLEAN-like algorithms \cite{CLEAN}, may fail in retrieving the required precision \cite{Zechmeiste,Pires},  mainly because these approaches rely on a white noise assumption. In order to increase the precision of the fit, the noise correlation matrix must also be estimated. A general approach is to maximize the likelihood function with respect to both regression parameters ($\delta$ in our problem) and the noise correlation matrix. Such an approach can use the Expectation-Minimization (EM) procedure like MAPES algorithms \cite{wang2005nonparametric}.  However, their convergence may be very slow, especially for large data samples like in the MICROSCOPE case (about $10^{6}$ points).  More recent works also use least squares iterative adaptive approaches (IAA) to estimate harmonic and noise parameters iteratively  \cite{stoica2009missing}, but require to store and invert correlation matrices, which is computationally expensive with an observation vector of $10^{6}$ entries. Likewise, the authors of the last two techniques do not present applications with colored noise. Some methods are already implemented to extract unknown colored spectral densities, especially in the domain of gravitational waves detection (see for example \cite{Roever11,PhysRevD.90.042003,Littenberg09,BayesLine}), but they do not tackle the problem of gapped time series. A suitable method is thus developed to estimate the EP parameter in case of missing data.

Another type of algorithms referred to as ``inpainting'' techniques is based on a sparsity-prior to fill the gaps \cite{Donoho,Elad}. 
Their adaptation to general noise spectra is currently studied in the MICROSCOPE team (Berg\'{e} \textit{et al.}, in prep.). We rather focus here on an approach that avoids filling the gaps.

We develop a method with two successive objectives. The first one is to reach the order of magnitude of the original (i.e. complete data) uncertainty in the estimation of the amplitudes of the deterministic components we are looking for. 
The second objective is to theoretically quantify the improvement on the variance of the estimator, using an approach that does not require to fill in the data gaps. 

Our technique is based on the estimation of the noise spectrum by using a high-order autoregressive (AR) model. The result is used to weight the data through an orthogonalization of the covariance matrix. This leads to an approximation of the best estimator in the sense of the variance, also referred to as the Best Linear Unbiased Estimator (BLUE) which is also the Generalized Least Squares (GLS) estimator in a linear regression context. The main idea in the proposed approach is to separately estimate the noise coefficients and the regression parameters instead of jointly estimating all the parameters. This is done in an iterative procedure that avoids the use of non-linear optimization algorithms. 

The proposed approach, that we dub ``Kalman-AR Model Analysis'' or ``KARMA'' for short, is divided in three steps. The first step consists in estimating the AR parameters describing the noise. This is done by using Burg's algorithm adapted to discontinuous data \cite{de2000burg}. The second step is carried out via a Kalman filter algorithm based on the AR model that allows us to compute the weights, as shown by \citet{jones1980maximum}. In the third step we finally compute an approximation of the Generalized Least Squares (GLS) estimator of the regression parameters, in a way similar to maximum likelihood computation methods applied to regression models \cite{kohn1985efficient,gomez1994estimation}. These steps can be reproduced to converge to the maximum likelihood estimator (MLE) of the parameters.

In this paper, we first analyze the effect of the missing data pattern on the estimation uncertainty (section \ref{section:impact}). We then describe the KARMA method (section \ref{section:method}) and we present a way of evaluating its performance, allowing us to give a criteria for the detection of the searched signal (section  \ref{section:detection}). Finally, after a brief description of the mission context, we apply this technique to MICROSCOPE simulated time series, in particular to data samples generated with the mission and instrument simulator (section \ref{section:results}). In section \ref{section:conclusion} we discuss the results.

\section{\label{section:impact} Impact of missing data}

Although we apply our study to the MICROSCOPE data analysis, it can be viewed as a general regression problem. The measurement equation can be summarized as follows:
\begin{equation}
\bm{\gamma} = \delta \bm{s}_{\rm{EP}} + \sum_{i} \alpha_{i} \bm{s}_{p,i} + \bm{z},
\label{eq:measurement}
\end{equation}
where $\bm{\gamma}$ is the $N$-points complete measurement vector defined as $\bm{\gamma} = \begin{pmatrix} \gamma_{0} & \hdots & \gamma_{N-1} \end{pmatrix}^{T}$, and $\delta$ and $\bm{s}_{\rm{EP}}$ are respectively the parameter and the signal of interest (the EP parameter and the EP violation signal for our purpose).  

The second term accounts for possible perturbations, whose amplitudes $\alpha_{i}$ should also be estimated to reject any bias.

The third term is the residual noise vector $\bm{z}$ assumed to be a zero-mean Gaussian random vector. The main objective is the estimation of  $\delta$, for which the square root of the one-sided noise power spectral density (PSD) at EP frequency must be $1.4 \times 10^{-12}$ ms\up{-2}$/\sqrt{\rm{Hz}}$ \cite{Touboul}. 

The presence of missing or corrupted data in the time series is identified by a mask vector $\bm{w}$ which is equal to 1 when the data is available and 0 otherwise, regardless of the nature of the gap. The observed signal is thus the vector $\bm{y}$ with entries $y_{n} = w_{n}\gamma_{n}$. We assume that the loss of data arises before any possible filtering.

\subsection{Impact on the PSD}
We briefly derive the impact of the observation window $\bm{w}$ on the PSD of a pure stationary random signal. Thus in this section we assume $\bm{\gamma} = \bm{z}$. The real signal $\gamma$ is regularly sampled at a frequency $f_{s}$ so that $\gamma_{n} = \gamma(n/f_{s})$.

For a stationary discrete parameter process, the autocovariance function is defined as: 
\begin{equation}
R_{y}(k) \equiv \mathrm{E}[y_{n}y_{n+k}] - \mathrm{E}[y_{n}]\mathrm{E}[y_{n+k}].
\label{eq:autocov}
\end{equation}
Then the PSD is the Discrete-Time Fourier Transform (DTFT) of the autocovariance \cite{priestley1982spectral}:
\begin{equation}
S_{y}(f) = \frac{1}{f_{s}} \sum_{k=-\infty}^{+\infty} R_{y}(k) e^{-2j \pi k f / f_{s} }.
\label{eq:PSD}
\end{equation}
In the case of the masked noise $y_{n} = w_{n}z_{n}$, Eq. (\ref{eq:autocov}) gives:
\begin{equation}
R_{y}(k) = \mathrm{E}[w_{n}z_{n}w_{n+k}z_{n+k}] - \mathrm{E}[w_{n}z_{n}] \mathrm{E}[w_{n+k}z_{n+k}]. \nonumber
\label{eq:autocov_derivation1}
\end{equation}
We assume that the underlying process in $z$ is independent of the window $w$, and that $w$ is a stationary process, so that one can write:
\begin{eqnarray}
R_{y}(k) & = & \mathrm{E}[z_{n}z_{n+k}] \mathrm{E}[w_{n}w_{n+k}] - \mu_{z}^{2} \mu_{w}^{2} \nonumber \\
 & = & \left(R_{z}(k) + \mu_{z}^{2} \right) \left( R_{w}(k)  + \mu_{w}^{2} \right) - \mu_{z}^{2} \mu_{w}^{2}, 
\label{eq:autocov_derivation2}
\end{eqnarray}
where for any random variables $x$ we note $\mu_{x}$ its expectation.

Assuming that $z_{n}$ is a zero-mean process ($\mu_{z}=0$), the PSD of the windowed signal is obtained by taking the DFT of Eq. (\ref{eq:autocov_derivation2}):
\begin{equation}
S_{y}(f)   =  \mu_{w}^{2} S_{z}(f) + \left[ S_{w} \ast S_{z} \right](f),
\label{eq:observed_PSD}
\end{equation}
where $\ast$ is the convolution operator. 

The first term can be viewed as a loss of power due to the missing data and the second term accounts for the frequency leakage. In the case of uniform random gaps, one shows (see appendix \ref{annex:random_case}) that $\mu_{w}$ is equal to the probability to have a gap at a given time. Then $S_{w}(f)$ is a constant, and the leakage term is proportional to the mean power. Therefore, the noise will increase significantly at frequencies where the leakage term is dominant.

\begin{figure}[!h]
\centering
  \includegraphics[width=0.5\textwidth]{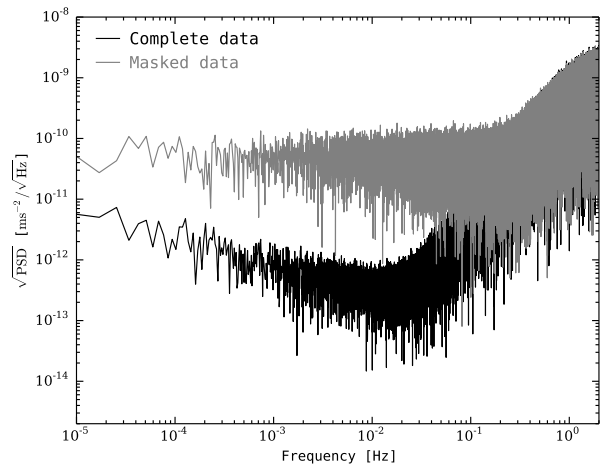}
\caption{\label{fig:1}Periodogram of original (black) and incomplete (grey) time series with 0.5 second data gaps randomly distributed in a 20 orbits session. The simulation is done for 260 random gaps per orbit.}
\end{figure}

As an illustration, a simulation of the MICROSCOPE instrumental noise alone is presented in Fig. \ref{fig:1}. The noise is generated using an approximate PSD model, taking into account thermal sensitivities at lower frequencies, position sensor noise at higher frequencies, random noise of the pick-up circuitry and the frequency response of the control loop:

\begin{equation}
2S_{z}(f) = \sigma_{z}^{2} \left( 1 + \left( \frac{f}{f_{1}} \right)^{-1} +  \left( \frac{f}{f_{2}}\right)^{4} \right) \cdot \left| H_{cl}(f)\right|^{2}
\label{eq:PSD_model}
\end{equation}
with $\sigma_{z} = 1.4 \times 10^{-13}$ ms\up{-2}$/\sqrt{\rm{Hz}}$, $f_{1} = 8.1 \times 10^{-2}$ Hz and $f_{2} = 1.3 \times 10^{-2}$ Hz. $H_{cl}$ is the transfer function of the closed control loop of the accelerometer. It has almost a unit gain for all frequencies under 1 Hz, and induces a slight inflection in higher frequencies.
The factor of $2$ accounts for the fact that $S_{z}(f)$ is the two-sided PSD.
The data is sampled at a frequency $f_{s} = 4$ Hz on a duration $T = 1.4$ days corresponding to 20 satellite orbits with $N_{g} = 5200$ gaps of the same length (0.5 seconds), randomly distributed over the time series. We observe a transfer of power from high frequencies to low frequencies, increasing the apparent noise around $f_{\rm{EP}} = 9.4 \times 10^{-4}$ Hz by two orders of magnitude.

\subsection{Impact on the least squares estimate}
We now demonstrate that the observed increase of the noise is not a simple artifact of the Fourier representation but directly impacts the estimation uncertainty in a least squares fitting approach. We assume that the analyzed signal is the sum of a harmonic component $\bm{s}_{\rm{EP}}$ at frequency $f_{\rm EP}$ and a correlated Gaussian random noise $\bm{z}$. For the sake of simplicity, we ignore the presence of possible deterministic perturbations, therefore the signals $s_{p,i}$'s in Eq. (\ref{eq:measurement}) are all zero. The signal is still sampled at frequency $f_{s}$ on $N$ data points. Thus the signal reads:
\begin{equation}
\bm{\gamma} = \delta \bm{s}_{\rm{EP}} + \bm{z}.
\label{eq:model}
\end{equation}

We define the window matrix as the diagonal matrix formed by the window vector: $W =  \mathrm{diag} \begin{pmatrix} w_{0} & \hdots & w_{N-1} \end{pmatrix}$. We aim at calculating the variance of the OLS estimate that only uses the available data (at times for which $w_{n}=1$). In the least squares formalism, this is equivalent to studying the windowed vector $\bm{y} = W \bm{\gamma}$. We also define the model matrix $A$. Although it can take a general form including various signals, we assume here that it contains the EP signal model only such that $A = \bm{s}_{\rm EP}$. We also define the masked model matrix $A_{w} = W A$. 
The usual OLS formulas give the following parameter estimate:
\begin{equation}
	 \label{equ:OLS_estimator}
   \hat{\delta} = ({A_{w}}^{\dag}A_{w})^{-1} \cdot {A_{w}}^{\dag} \bm{y},
\end{equation}
as well as its variance:
 \begin{eqnarray}
	 \label{equ:OLS_covariance}
   \mathrm{Var}(\hat{\delta}) & = & K^{-1}{A_{w}}^{\dag} \Sigma_{y} A_{w} K^{-1},
\end{eqnarray}
where we defined $K = {A_{w}}^{\dag}A_{w}$ and $\Sigma_{y} = W \Sigma_{z} W^{\dag}$ with $\Sigma_{z} = \mathrm{E}[\bm{z}\bm{z}^{\dag}]$, the covariance matrix of the noise vector. Here $\dag$ denotes the hermitian adjoint. 
As a result, the noise correlation seen by the estimator is $\Sigma_{y}$ instead of $\Sigma_{z}$ in the complete case. 

In the case of a stationary Gaussian random noise the estimator covariance can be diagonalized in the Fourier space:
\begin{equation}
\Sigma_{y} = \frac{f_{s}}{N} M^{\dag} D M,
\label{eq:PSD_covariance}
\end{equation}
where $D$ is the diagonal matrix formed by the two-sided discrete PSD: $D = \mathrm{diag} \begin{pmatrix} \hat{S}_{0} & \hdots & \hat{S}_{N-1} \end{pmatrix}^{T}$ and $M$ is the Discrete Fourier Transform (DFT) matrix with coefficients: $M_{kl} = \exp\left(-\frac{2 i \pi k l}{N}\right) $. The discrete spectrum is defined as the expectation of the periodogram. It can be seen as an approximation of the real PSD \cite{priestley1982spectral}:
\begin{equation}
\hat{S}_{k} \equiv \frac{1}{f_{s}} \sum_{n=-(N-1)}^{N-1} \left(1 - \frac{|n|}{N}\right) R_{y}(n) e^{-2j \pi \frac{nk}{N} }.
\label{eq:discrete_spectrum}
\end{equation}

\begin{figure}[!h]
\centering
  \includegraphics[width=0.5\textwidth]{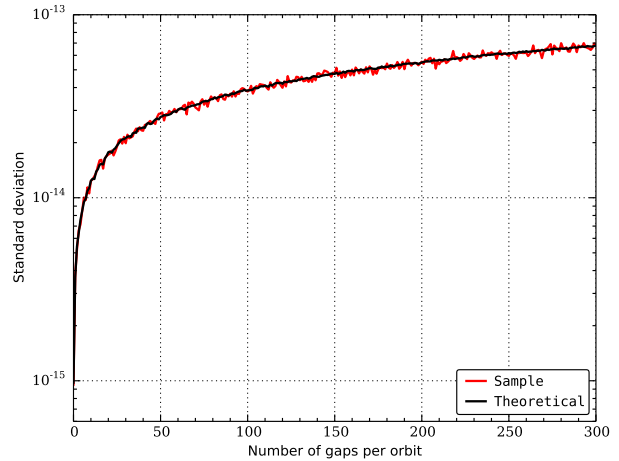}
\caption{\label{fig:2}Theoretical (black) and sample (red) standard deviations of the original least squares estimate of the EP parameter as a function of the number of gaps per orbits. All gaps have the same duration of 0.5 second and are randomly distributed over a 20 orbits session.}
\end{figure}

This diagonalization thus links the estimator variance and the PSD of the windowed noise. By developing Eq. (\ref{equ:OLS_covariance}) we show (see appendix \ref{annex:OLS_variance} for more details) that in the case of a harmonic model such as $s_{\rm{EP},n} = \gamma_{\rm EP} \sin( 2 \pi n f_{\rm EP} / f_{s} )$, for sufficiently large $N$, the estimator variance is approximately equal to:
\begin{equation}
\mathrm{Var}(\hat{\delta}) \approx \frac{2f_{s} N S_{y}(f_{\rm{EP}}) }{N_{o}^{2} \gamma_{\rm{EP}}^{2}},
\label{eq:masked_variance}
\end{equation} 
where $S_{y}$ is given by equation (\ref{eq:observed_PSD}), $N_{o} = N - N_{g}$ is the number of observed data and $\gamma_{\rm{EP}}$ is the amplitude of the model, which is the gravitational acceleration in our case. As a result, in presence of missing data, the estimation variance increases proportionally to the leakage term in Eq. (\ref{eq:observed_PSD}). To quantify the increase of the uncertainty, we plot the standard deviation of the estimator as a function of the number of data gaps per orbit in Fig. \ref{fig:2}, in the case of short random gaps of fixed length (0.5 second) uniformly distributed over the time series (the effect of the size and the number of gaps is discussed in Berg\'e \textit{et al.}, in prep.). The theoretical standard deviation (black curve) is obtained using Eq. (\ref{equ:OLS_covariance}). In order to check the correctness of the distribution, we also plot the sample standard deviation of 400 estimates (red curve) corresponding to different realizations of the noise vector $\bm{z}$. This shows that the uncertainty grows by one order of magnitude from 10 gaps per orbit only, which represents a data loss of 0.04 \%. This is not acceptable with respect to the performance objectives of the mission. Therefore an alternative estimation method needs to be implemented.

\section{\label{section:method} Kalman-AR Model Analysis (KARMA)}

The poor performance of the OLS estimator is due to the fact that its variance is not minimal. To minimize the variance, the Best Linear Unbiased Estimator (BLUE) is needed, which takes the form of a Generalized Least Squares (GLS) estimator in linear regression problems. In case of missing data, it reads:
\begin{equation}
\hat{\bm{\beta}} = ({A_{o}}^{\dag}{\Sigma_{o}}^{-1}A_{o})^{-1} \cdot {A_{o}}^{\dag}{\Sigma_{o}}^{-1}\bm{y_{o}},
\label{eq:GLS}
\end{equation}
where $\bm{\beta}$ is the $q \times 1$ vector of parameters to be estimated. The observation vector $\bm{y_{o}} \equiv \begin{pmatrix} \gamma_{n_{0}} & \hdots & \gamma_{n_{N_{o}-1}}  \end{pmatrix}^{T}$ gathers the available data only, that is, $n_{0}, \hdots, n_{N_{o}-1}$ are the time indexes corresponding to the observed data. Similarly, $A_{o}$ is the model matrix where we have kept only rows corresponding to observed data. Here $A_{o}$ is assumed to be general, of size $N \times q$. $\Sigma_{o}$ is the covariance matrix of the observed noise vector $\bm{z_{o}}$ and admits a Cholesky decomposition such that  $\Sigma_{o} = L_{o} {L_{o}}^{\dag}$ where $L_{o}$ is a lower triangular matrix.

The difficulty here is to estimate the noise covariance matrix $\Sigma_{o}$ in spite of the missing data. The method that we propose consists in calculating an approximation of the GLS estimator by postulating an autoregressive (AR) model for the noise. This is done in three steps which are detailed below: estimation of AR parameters (step 1), calculation of the whitened vectors ${L_{o}}^{-1}\bm{y_{o}}$ and ${L_{o}}^{-1}A_{o}$ (step 2), calculation of the estimate $\hat{\bm{\beta}}$ (step 3). The process may be iterated if necessary.

\subsection{Step 1: AR parameters estimation}
\label{subsection:AR_estimation}
The first step is to estimate the noise characteristics encapsulated in the covariance matrix $\Sigma_{z}$. To do so, we assume that the noise process can be described by an autoregressive model of some order $p$ to be determined, verifying the following relation at all times $n$:
\begin{equation}
z_{n} + a_{1}z_{n-1} + ... + a_{p}z_{n-p} = \epsilon_{n},
\label{eq:AR_model}
\end{equation}
where $a_{1}, \hdots, a_{p}$ are the AR coefficients and $\bm{\epsilon}$ is a zero-mean white Gaussian random field of variance $\sigma^{2}$. Note that this is equivalent to approximating the noise PSD with a rational function, the numerator being a polynomial of degree $p$ in $\exp(-2 i \pi f/f_{s})$ such as:
\begin{equation}
\hat{S}_{z}(f) = \frac{\sigma^{2}/f_{s}}{\left| 1 + a_{1}e^{-2 i \pi f/f_{s}} + ... + a_{p}e^{-2 i \pi p f/f_{s}} \right|^{2}}.
\label{eq:AR_spectrum}
\end{equation}

The choice of this model is motivated by the following arguments. Firstly the use of a parametric model consistently reduces the number of noise parameters to estimate ($p$ instead of $N$), and therefrom the computational cost. Secondly choosing an AR model rather than a more general class such as autoregressive-moving average models (ARMA) allows us to easily estimate the parameters from the discontinuous data, while ARMA models usually involve computationally expensive optimization procedures, or direct estimation of the autocovariance function which is not accurate when data are missing. Furthermore any moving-average model can be approximated by a high order AR model as discussed by \citet{Durbin}.

The AR parameters $\bm{\theta} = (a_{i}, \sigma^{2})$ are estimated thanks to Burg's algorithm adapted to the missing data case \cite{de2000burg}. This technique relies on the minimization of forward and backward residuals of the model ($\ref{eq:AR_model}$) through a recursive procedure that increases the order $k$ of the AR model at each step, until $k$ reaches $p$. This algorithm takes advantage of all segments of available data. For a given order $k$, only the segments of size $N_{s} > k$ can be used for the estimation. Note that the proper AR order must be previously determined according to some criteria such as Akaike's, as discussed later.

For the first iteration, the AR estimation is performed on the residuals of the OLS estimation $\bm{\hat{z}_{o}} = \bm{y_{o}} - A_{o} \hat{\bm{\beta}}_{\rm{OLS}}$ instead of $\bm{y_{o}}$, where $\hat{\bm{\beta}}_{\rm{OLS}}$ is the result of the simple estimate given by Eq. (\ref{equ:OLS_estimator}). This reduces the disturbance of deterministic components onto the estimation of the noise parameters.

\subsection{Step 2: computation of the weighted vectors with the Kalman filter}
\label{subsection:weights}
The determination of the AR parameters gives access to the noise autocovariance function. The aim of this step is to use this result to calculate the weighted observation vector $L_{o}^{-1}\bm{y_{o}}$ and weighted model matrix $L_{o}^{-1}A_{o}$ involved in the expression of the estimator (\ref{eq:GLS}). The matrix $L_{o}$ indirectly depends on the AR parameters via the autocovariance function, since:
\begin{equation}
\Sigma_{o}(m,l) = R_{z}\left( |n_{m}-n_{l}| \right) \text{  } \forall (m,l) \in \llbracket 0 , N_{o}-1\rrbracket ^{2},
\label{eq:covariance_elements}
\end{equation}
where the autocovariance function $R_{z}$ is estimated by taking the inverse Fourier transform of Eq. (\ref{eq:AR_spectrum}).

Unlike the case of complete stationary random series, the observed data in a missing pattern do not have a circulant nor Toeplitz correlation matrix, because the $n_{i}$'s are not regularly arranged. Therefore the matrix $\Sigma_{o}$ cannot be inverted by efficient techniques such as Levinson or FFT algorithms. If the data sample is large (like in the MICROSCOPE case where typically $N \sim 10^{6}$), this creates memory difficulties to store such a matrix. That is why we present a way of avoiding the direct inversion using a Kalman algorithm to compute the weighted data.

The relationship between GLS and Kalman filtering is explained as follows. Following the notation of \citet{gomez1994estimation}, an AR process can be described by the state-space representation: 
\begin{eqnarray}
	\bm{x}(n) & = & F \bm{x}(n-1) + G\epsilon(n), \\
	    z_{n} & = & H^{T} \bm{x}(n).
\label{eq:state}
\end{eqnarray}
The above equations are the state equation and the observation equation of the Kalman Filter. $\bm{x}(n)$ is the state vector at time $n$, defined by:
\begin{eqnarray*}
	\bm{x}(n) \equiv \begin{pmatrix} z_{n} & z_{n+1|n} & \ldots & z_{n+p-1|n} \end{pmatrix}^{T},  
\end{eqnarray*}
where $z_{n+k|n}$ is the conditional expectation of $z_{n+k}$ given the observations before time $n$. 
$H$ is the matrix linking the state vector to the observations, and simply reads $H = \begin{pmatrix} 1 & 0 & \hdots & 0 \end{pmatrix}^{T}$. The model matrix $F$ and the model noise vector $G$ are calculated from the AR parameters and are defined in appendix \ref{annex:state_space}.

The Kalman filter aims at calculating the \textit{a priori} estimate of the state vector along with its variance at each time $n$ given all the observations until time $n-1$, that is:
\begin{eqnarray}
	         z_{n|n-1} & \equiv &  \mathrm{E}[z_{n}|z_{0},z_{1},\hdots,z_{n-1}], \\
	\sigma^{2}_{n|n-1} & \equiv & \mathrm{Var}[z_{n}|z_{0},z_{1},\hdots,z_{n-1}].
\end{eqnarray}

We define the normalized innovation vector $\bm{e}$ whose elements are calculated with the Kalman residuals and their standard errors:
\begin{equation}
e_{n} \equiv \left(z_{n} - z_{n|n-1} \right) / \sigma_{n|n-1}.
\label{eq:innovation}
\end{equation}
Since $z_{n|n-1}$ is actually the projection of $z_{n}$ onto the subspace generated by $\begin{pmatrix} z_{0} & \ldots & z_{n-1} \end{pmatrix} $, Eq. (\ref{eq:innovation}) is equivalent to a Gram-Schmidt orthogonalization procedure. As a result, the $e_{n}$'s are uncorrelated. In addition, $z_{n|n-1}$ is a linear combination of $z_{i} \text{, } i < n$, thus the normalized innovation vector can be expressed as:
\begin{equation}
\bm{e} = T \bm{z},
\label{eq:relation_e_z}
\end{equation}
where $T$ is a lower triangular matrix with diagonal elements equal to one.
If we calculate the autocovariance of Eq. (\ref{eq:relation_e_z}) we find that 
\begin{equation}
\mathrm{Cov}[\bm{e}] = T \Sigma_{z} T^{\dag} \Rightarrow \Sigma_{z} = ( T T^{\dag} )^{-1},
\label{eq:e_covariance}
\end{equation}
where the implication is based on the fact that $\mathrm{Cov}[\bm{e}]$ is equal to the identity matrix. This last equation shows that the matrix $T$ is equal or proportional to the inverse of the Cholesky decomposition $L^{-1}$ of the covariance matrix $\Sigma_{z}$. However, in our problem this is not exactly true. The derived equalities are only valid if the random data truly follows the AR process, which is not the case in our approach since the AR model is just an approximation of the real underlying random process. We thus assume that the Kalman output $\bm{e}$ is only approximately equal to $L^{-1}\bm{z}$.

If data are missing, the classic Kalman procedure must be slightly modified to properly deal with missing data, as explained by \citet{jones1980maximum}, but the components of the normalized innovation vector $\bm{e}$ corresponding to missing data are ignored in the estimation at step 3.

\subsection{\label{subsection:GLS} Step 3: computation of the GLS estimate}

In the previous paragraph we showed how to perform a quasi orthogonalization of the observation vector, which is exactly what is needed to compute an approximate version of the Generalized Least Squares (GLS) estimate. 

The estimator in Eq. (\ref{eq:GLS}) can be rewritten:
\begin{equation}
\hat{\bm{\beta}} = ({E_{o}}^{\dag} E_{o})^{-1} \cdot {E_{o}}^{\dag} \bm{e}_{o},
\label{eq:GLS_Kalman}
\end{equation}
where, with obvious notation, we denote the normalized innovation vectors $\bm{e_{o}} = T_{o}\bm{y_{o}}$ and $E_{o} = T_{o} A_{o}$, calculated with the outputs of the Kalman filter algorithm, respectively applied to the observed signal and to each columns of the model matrix. Both vectors are obtained by keeping elements corresponding to observed data only. The Kalman algorithm is thus used here as a device to compute the weighted vectors involved in the GLS.

\section{\label{section:detection} Theoretical uncertainty and detection issues}

This is of key interest to be able to assess the statistical uncertainty of a given estimation, especially in a context where the experiment cannot be reproduced a large number of times. In this section we present a tool to quantify the uncertainty of the regression result and to give a confidence threshold for the detection of the signal of interest. To achieve this goal, the estimator variance matrix must be estimated.

The correlation matrix can be approximated under the assumption that the AR model is a good approximation of the real noise correlations. This hypothesis is equivalent to assuming that the estimator has minimal variance (\textit{i.e.} that the estimator is the BLUE). Let $C$ be the covariance matrix of the estimator $\hat{\beta}$. Then Eq. (\ref{equ:OLS_covariance}) gives, by replacing $W$ by $T_{o}$ and $A$ by $A_{o}$:
\begin{equation}
\hat{C} \approx \sigma_{0}^{2} \left( {E_{o}}^{\dag} E_{o} \right)^{-1},
\label{eq:Sigma_approx}
\end{equation}
where $\sigma_{0}$ accounts for the fact that the covariance is known up to a proportionality constant. For an unbiased estimator (i.e. the model matrix $A_{o}$ describes all the deterministic components of the signal) this can be estimated by:
\begin{equation}
\hat{\sigma}_{0}^{2} =  \frac{{\bm{\hat{e}_{z}}}^{\dag}\bm{\hat{e}_{z}}}{N_{o}-q},
\label{eq:sigma0}
\end{equation}
where $\bm{\hat{e}_{z}}$ is the vector of weighted residuals defined by $\bm{\hat{e}_{z}} \equiv \bm{e_{o}} - E_{o} \hat{\bm{\beta}}$. 
The statistic to be considered is:
\begin{equation}
Z_{k} \equiv \frac{\hat{\beta}_{k}}{\sqrt{\hat{C}_{k,k}}},
\label{eq:statistics}
\end{equation} 
where $k$ is the index corresponding to the parameter of interest in the vector $\bm{\beta}$. For our application $\beta_{k}$ is the EP parameter $\delta$.
Here we assume that the underlying process is Gaussian, which is reasonable in the case of the MICROSCOPE instrumental noise.
Then under the assumption that there is no violation signal (hypothesis $H_{0}$), $Z$ approximately follows a Normal law with mean zero and unit variance. 
A detection threshold with a $(1-\alpha)$-confidence level is given by imposing that the probability to observe a value above the threshold, under $H_{0}$, must be lower than $\alpha$. This gives the threshold $z = \Phi^{-1}\left(1-\frac{\alpha}{2}\right)$, where $\Phi$ is the Normal Cumulative Distribution Function (CDF). Therefore if $|Z|$ is above the threshold, then a signal is detected with a confidence of $100(1-\alpha)\%$. Conversely, for a given estimation of the EP signal, the violation can be claimed at a $100 ( 2 \Phi(Z) - 1) \%$ confidence level. Typically, a 99\% confidence test requires $z = 2.56$.

\section{\label{section:results}Application to simulated data of the MICROSCOPE mission}

\subsection{The MICROSCOPE experiment}
The Weak Equivalence Principle (WEP) is at the basis of General Relativity. Its concrete manifestation is the Universality of Free Fall, stating that a body in a gravitational potential is accelerated independently of its mass and internal composition. Current efforts to build new unification theories may call this principle into question \cite{Damour}, postulating the existence of additional fundamental interactions. To provide an experimental discrimination of these theories, the goal of the MICROSCOPE space mission is to test the WEP within a precision of about $10^{-15}$ never reached by previous ground-based experiments \cite{wagner2012torsion,Will}. This space-borne test takes advantage of the duration of the fall by integrating the data over several orbits.

The mission payload is an ensemble of two electrostatic differential accelerometers composed by a cage containing two cylindrical and co-axial test-masses (TM). One accelerometer is devoted to the EP test, while the other serves as a reference. In the first accelerometer, the two TM have different compositions: one is made of Platinum Rhodium alloy (PtRh) and the other of Titanium alloy (TA6V) \cite{Hardy20131634}. In the second accelerometer the TM are both made with PtRh. The masses, whose potential is kept constant via a thin gold wire, are servo-controlled by a set of electrodes to follow the same trajectory. The MICROSCOPE science signal is the difference between the accelerations applied to the two TM, which are deduced from the applied electrostatic forces needed to maintain them relatively motionless at the center of the cage. A drag-free system ensures that the measured common acceleration, \textit{i.e.} the mean acceleration of the two TM, is nullified. A violation of the WEP would result in a difference between the two measured accelerations.

The violation signal is expected to be periodic with a frequency $f_{\rm{EP}}$ because of the projection of the gravitational acceleration onto the science axis of the instrument during the orbital trajectory. For a satellite inertial pointing session, $f_{\rm{EP}}$ is equal to the orbital frequency. For a slowly rotating satellite in the orbital plane, this is equal to the sum of the orbital frequency and the satellite spin frequency. The duration of each session is chosen in order to reach a standard deviation error of about $10^{-15}$ on the EP parameter $\delta$, which is almost equal to the E\"{o}tv\"{o}s parameter.
The inertial and spin sessions last respectively 120 and 20 orbits. The specificity of the data samples to be analyzed in the MICROSCOPE mission is that the signal of interest has a low signal-to-noise ratio (SNR) that lies at low frequencies ($10^{-4}$ - $10^{-3}$ Hz) in a time series with a broad frequency range ($10^{-5}$ - $2$ Hz), blurred by a colored noise containing most of its power in higher frequencies (above $10^{-1}$ Hz). In addition, long time series must be analyzed to achieve a sufficient SNR, including about $5 \times 10^{5}$ data points for a spin session.

\subsection{Considered data sets}
We apply the KARMA method to a time series simulated with a mission simulator. The simulation output is the differential acceleration vector $\bm{\gamma}$ equal to the acceleration difference between the two masses. This time series is sampled at $f_{s} = 4$ Hz and lasts 20 orbits. This corresponds to a spin session, for which the orbital frequency is equal to $1.7 \times 10^{-4}$ Hz and the spin frequency is $7.7 \times 10^{-4}$ Hz. The EP frequency is then equal to the sum $f_{\rm{EP}} = 9.4 \times 10^{-4}$ Hz.

In addition to the signal of interest, other perturbations are present in the measurement as indicated in Eq. (\ref{eq:measurement}). They are mainly due to gradient terms between the center of mass of the two TM, the relative motion of the TM, and coupling with the common mode because of instrument defects. During the experiment, the instrument or the satellite undergoes excitations that favor the SNR to measure their amplitudes $\alpha_{i}$. The corresponding accelerations $\bm{s}_{p,i}$ are either modeled or measured, such that the perturbations can be subtracted from Eq. (\ref{eq:measurement}). 

Nevertheless, in this simulation we allow for the presence of gradient perturbations. They come from the slight off-centering of the test-masses, leading to gravity and inertia gradient terms. In the simulation we assume that the TM are off-centered by 20 microns along the $x$ and $z$-axis which are in the orbital plane. Note that although an off-centering along the $y$-axis can also exist, it is estimated by means of dedicated calibration sessions and corrected numerically before the EP estimation. The EP parameter is simulated at a level of $3 \times 10^{-15}$. Thus the regression model $A$ contains the true acceleration signal $g_{x}(t)$, to which we add the two perturbations modeled with our knowledge of gravity and inertia gradients. The noise added to the data is generated from the PSD model given by Eq. (\ref{eq:PSD_model}). The signal model reads:
	\begin{equation}
	\gamma(t) =  \delta g_{x}(t) + \Delta_{x} T_{xx}(t) + \Delta_{z} T_{xz}(t) + z(t),
	\label{eq:case2_model}
	\end{equation}
where we have noted $g_{x}$ the gravitational acceleration projected onto the $x$-axis, $T_{ij}$ the components of the gradient tensor, and $\Delta_{i}$ the off-centerings. Thus in this case there are three regression parameters: $\delta$, $\Delta_{x}$ and $\Delta_{z}$.

\begin{figure}[!h]

\centering
  \includegraphics[width=0.5\textwidth]{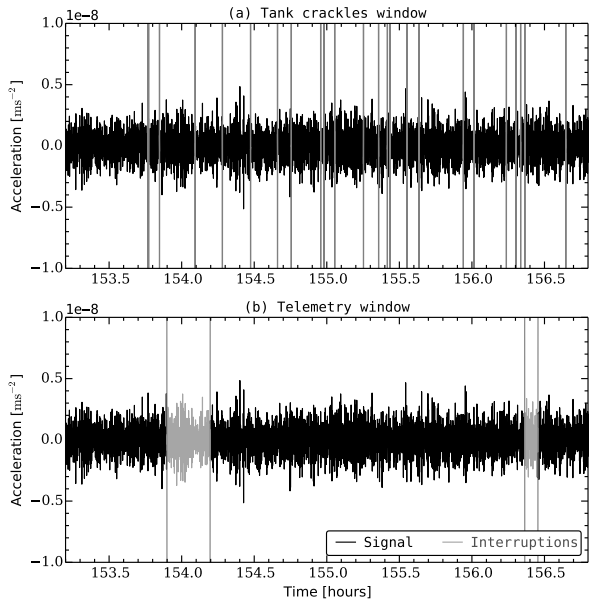}
\caption{\label{fig:3}Fraction of the temporal series (black) with the interruption times represented by the grey vertical lines for the two windows $\bm{w}_{a}$ (top) and $\bm{w}_{b}$ (bottom).}
\end{figure}

We consider two types of gap pattern. The first one is a ``tank crackle type'' window $\bm{w}_{a}$ that is generated so that all gaps are of equal duration (0.5 second) and their positions are randomly distributed on the sample (uniform distribution with 260 gaps per orbit). The second one is a ``telemetry losses type'' window $\bm{w}_{b}$ where the gaps durations are drawn from a distribution similar to the telemetry thread of the PICARD mission \cite{PICARD}, with a standard duration of one minute. Their positions are distributed in the same way as for the first window. Each window represents the same fraction of missing data, of about 2\%. Thus window $\bm{w}_{a}$ comprises more gaps than window $\bm{w}_{b}$ (larger $N_{g}$) but gaps are shorter in average. To illustrate this, we plot in Fig. \ref{fig:3} an extract of the time series where the data interruptions of each window are identified by vertical grey bars.

We apply the KARMA method and compare the result to the Ordinary Least Squares estimate with missing data to assess the improvement. We also compare the result to the reference given by the OLS estimator in the case without gaps.

\subsection{PSD estimate from the AR fit}
Before starting the whole process, the order of the AR model must be chosen at step 1.
The choice of the order depends on the PSD of the noise affecting the measurement, and on the observation window. A way to properly choose the order $p$ is to minimize the Akaike's Information Criterion \cite{akaike1974new} defined as $\rm{AIC}(p) = 2p - 2\log\left(L_{\rm{max}}(p)\right)$, where $L_{\rm{max}}$ is the maximized log-likelihood. In the case of an AR model, this can be expressed in terms of the estimate of the AR residual variance $\hat{\sigma}^{2}$ which is directly computable from the residuals of the Burg's algorithm:
\begin{equation}
\rm{AIC}(p) = 2p + N_{o} \log( \hat{\sigma}^{2} ).
\label{eq:AIC}
\end{equation}
Applying Burg's algorithm to the residual series $\hat{\bm{z}}$ defined in section \ref{subsection:AR_estimation} with increasing order $k$ allows us to find the order that minimizes the AIC. 

In the MICROSCOPE case, $\rm{AIC}(p)$ is an asymptotically monotonic decreasing function. In this configuration one possibility is to choose the order $p$ from which there is no significant improvement in the AIC, \textit{i.e.} the minimum order where the AIC is close enough to the asymptote. For the studied noise, the AIC typically reaches a plateau from $p = 200$. 

\begin{figure*}
    \centering
    \begin{tabular}{c c}
      \multicolumn{2}{c}{\includegraphics[width=1.0\textwidth]{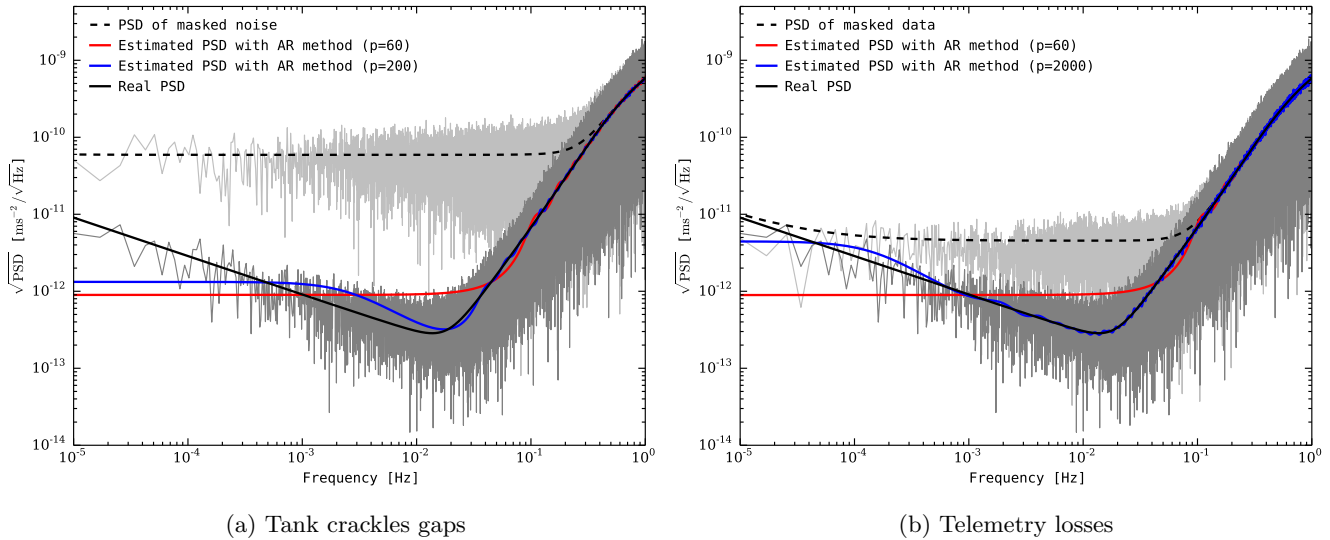}} \\
      {\hspace{3cm} (a) Tank crackles gaps} & {\hspace{2.5cm} (b) Telemetry losses}
    \end{tabular}
    \caption{\label{fig:4}PSD estimates of the noise in presence of missing data. The black dashed curve is an estimate of the masked data PSD [obtained using Eq. (\ref{eq:observed_PSD})], the black solid curve is the actual noise PSD, and the red and blue curves are the PSD estimates of the AR model obtained with Burg's algorithm. The periodograms of the regression residuals are also plotted for the complete (dark grey) and masked (light grey) cases. }
\end{figure*}

Nevertheless, in case of very frequent missing data (e.g. tank crackles), the variance of the AR coefficients estimates increases with the order, and so does the variance of the AIC. This is due to the decrease of the number of usable data segments (with length larger than $p$). This can lead to overestimating the optimal order $p$. To overcome this difficulty we can modify the AIC criterion as suggested by \citet{bos2002autoregressive} by introducing a penalty accounting for the increasing estimation variance. We choose to replace $N_{o}$ by $p \left( \sum_{i=1}^{p} \frac{1}{N_{i}} \right)^{-1} $ where $N_{i}$ is the number of usable segments to estimate the coefficient $a_{i}$. When applying this criterion to our simulation with window $\bm{w}_{a}$, we find an optimal order of $p=60$.

The process converges after 2 iterations, because the first estimate of the PSD is influenced by the high amplitude perturbations of the gradient terms: the main peak has an amplitude of $2.4 \times 10^{-11}$ ms\up{-2} at $2f_{\rm{EP}}$, and other peaks are present at $f_{orb}$ and $2f_{orb}$. In comparison, the EP violation corresponds to an amplitude of $1.2 \times 10^{-14}$ ms\up{-2}.

We plot in Fig. \ref{fig:4} the estimate of the PSD (red curve) obtained with the AR coefficients calculated by Burg's algorithm with the tank crackles (a) and telemetry (b) windows, along with the real PSD (black curve). The level of noise of the masked data is shown by the black dotted line. In addition to the selected order $p = 60$, we also show the AR spectrum estimate made with a larger order ($p=200$ with window $\bm{w}_{a}$, and $p = 2000$ with window $\bm{w}_{b}$) to illustrate the effect of $p$. In both cases, we see that the overall shape of the PSD is well described by the AR model, especially the $f^{4}$ slope. However, there is a bias which increases as the frequency decreases. The reasons why the AR model cannot accurately describe the low frequency PSD are two-fold:
\begin{enumerate}
	\item The order of the AR model is finite, and limited by the longest segment of consecutive available data (this is typically 700 for the window $\bm{w}_{a}$, and $50000$ for $\bm{w}_{b}$). Given that AR models cannot describe $1/f$ spectra with a finite number of parameters, a larger order is necessary to reduce the bias (and the bias is zero when $p$ tends to infinity).
	\item In the Burg estimation procedure, the larger the AR order, the larger the variance of the AR coefficients estimates, because there are fewer segments of corresponding lengths. This is why we do not choose the highest possible order, for which segments of corresponding length are rare.
\end{enumerate}
Since window $\bm{w}_{b}$ has more spaced and longer gaps than window $\bm{w}_{a}$, it allows for a higher possible AR order leading to a better restitution of the low frequency shape of the PSD, with a reasonable variance (see Fig. \ref{fig:4}). However we choose $p=60$ even in the case of window $\bm{w}_{b}$ for computational reasons, given that this is the high frequency restitution that matters for a parameter regression purpose, as we shall see in the next paragraph.

\subsection{Regression results}
The results of the linear regression are summarized in Table \ref{table:error_case2}, with $p=60$. In order to test the precision of our method, we have drawn 400 realizations of the noise and run our estimation algorithm for each of them, as well as the OLS estimator. The number of draws is chosen such that the error on the true value of the standard deviation of the EP parameter does not exceed $10^{-16}$ with a 99\% confidence.

\begin{table*}
\caption{\label{table:error_case2}Mean and standard deviations on the estimation of the parameters of interest using OLS and the KARMA method. In both cases we present (from left to right) the estimation average calculated on a sample of 400 estimates, the analytical standard deviation, and the sample standard deviation. For OLS, the analytical uncertainty $\sigma_{th}$ is given by Eq. (\ref{equ:OLS_covariance}), which is exact. For the KARMA method, $\hat{\sigma}_{AR}$ is the average of the uncertainties estimated for each draw with Eq. (\ref{eq:Sigma_approx}). } 
\begin{center}
\begin{ruledtabular}
\begin{tabular}{ c  c  c | c  c  c | c  c  c }
\multirow{2}{*}{}                    & \multicolumn{2}{c|}{}             & \multicolumn{3}{c|}{Ordinary Least Squares}    & \multicolumn{3}{c}{Kalman-AR Model Analysis}  \\ 
\hline
    {Window}                         & {Param.}      & {True}  & {$\hat{\mu}$} & {$\sigma_{th}$}& {$\hat{\sigma}$} & {$\hat{\mu}$} & {$\hat{\sigma}_{AR}$} & {$\hat{\sigma}$} \\      \hline
    \multirow{3}{*}{Complete data}   & $\delta$ [$10^{-15}$] & $3$     & $3.01$  & $0.96$   & $1.02$   & $2.98$ & $0.92$  & $0.96$\\
		                                 & $\Delta_{x}$ [$\mu$m] & $20$    & $20.0$  & $0.003$  & $0.005$  & $20.0$ & $0.004$ & $0.003$\\
		  															 & $\Delta_{z}$ [$\mu$m] & $20$    & $20.0$  & $0.003$  & $0.005$  & $20.0$ & $0.004$ & $0.003$\\
		
		\hline
    \multirow{3}{*}{Tank crackles}   & $\delta$ [$10^{-15}$] & $3$    & $8.82$ & $62.3$  &$65.2$  &  $2.98$    & $1.19$   & $1.14$ \\
		                                 & $\Delta_{x}$ [$\mu$m] & $20$   & $20.0$ & $0.290$ &$0.296$ &  $20.0$    & $0.006$  & $0.004$ \\
		  															 & $\Delta_{z}$ [$\mu$m] & $20$   & $20.0$ & $0.292$ &$0.314$ &  $20.0$    & $0.006$  & $0.005$ \\
		
		\hline
		
		\multirow{3}{*}{Telemetry}     & $\delta$  [$10^{-15}$] & $3$  & $3.15$ & $5.20$   & $5.07$   & $2.98$     & $0.93$    & $0.98$ \\ 
		                               & $\Delta_{x}$ [$\mu$m] & $20$  & $20.0$ & $0.021$  & $0.021$  & $20.0$     & $0.004$   & $0.003$ \\
		  													   & $\Delta_{z}$ [$\mu$m] & $20$  & $20.0$ & $0.024 $ & $0.024$  & $20.0$     & $0.004$   & $0.003$ \\		
\end{tabular}
\end{ruledtabular}
\end{center}
\end{table*}

The third column of the table indicates the true value of the parameters. Columns 4 to 6 show the performance of the OLS estimator: the sample average $\hat{\mu}$, the theoretical standard deviation $\sigma_{th}$ given by Eq. (\ref{equ:OLS_covariance}) and calculated with the real PSD, and the sample standard deviation of the 400 estimates. The last three columns show the results obtained with the KARMA method, and are detailed below.

The sample mean $\hat{\mu}$ of the estimates obtained with the KARMA method converges to the true value of the parameters (seventh column of table \ref{table:error_case2}), showing that the constructed estimator is unbiased. 

We also calculate the sample standard deviation of the EP parameter. For short and numerous gaps (tank crackles window) we find $\hat{\sigma} = 1.1 \times 10^{-15}$ with our method instead of $6.5 \times 10^{-14}$ with the OLS estimator. Thus our method enables us to divide the stochastic error by a factor 60 with respect to the OLS. 

For fewer and longer gaps (telemetry window), we find $\hat{\sigma} = 9.8 \times 10^{-16}$ instead of $5.1 \times 10^{-15}$ with the OLS. We notice that such a gap pattern has less impact on the estimation performance, because it leads to a lower frequency leakage as confirmed by Eqs. (\ref{eq:observation_probability2}) and (\ref{eq:observed_PSD_MICROSCOPE}) of appendix \ref{annex:random_case} (also see Berg\'e \textit{et al.}, in prep.).

These are satisfying results since the theoretical uncertainty of the OLS without any missing data is equal to $9.6 \times 10^{-16}$. The detection test is positive with a confidence greater than 99\% in both cases. 

The improvement is also significant for the other parameters. Even if they are already well estimated by the OLS, their uncertainty is reduced by almost two orders of magnitude for the tank crackles window. 

For each draw, we estimate the uncertainty $\hat{\sigma}_{AR}$ using the approximate formula (\ref{eq:Sigma_approx}). We then calculate the sample average of this estimate over the 400 draws, and record the results in the table. We find $1.2 \times 10^{-15}$ for window $\bf{w}_{a}$ and $9.3 \times 10^{-16}$ for window $\bf{w}_{b}$. This is close to the calculated sample standard deviation, meaning that when having only one realization at hand, one can estimate the error with an acceptable accuracy. The estimated error does not vary much from one estimation to another, and stays within an interval of $\pm 10^{-16}$ around the mean.

The estimate $\hat{\sigma}_{AR}$ of the real regression error may be biased, depending on the frequency of the estimated signal. This can be explained by Fig. \ref{fig:4}, where we observe that the PSD of the AR model is biased at low frequency. As a result, the lower the signal frequency, the larger the bias on the estimated variance. This is particularly true around zero, where the AR PSD is below the real one.
However, the overall shape of the real PSD is well captured by the AR model, which is enough to cancel the leakage due to the window and get a precision of $1 \times 10^{-15}$ for the EP estimation, in agreement with the mission requirement.

\section{\label{section:conclusion}Conclusion and discussion}

We have shown that the presence of gaps in time series affected by correlated noise has a strong impact on the classical Fourier analysis and on the precision of the ordinary least squares fits of harmonic signals. This is due to the frequency leakage of the noise power, which can increase the uncertainty of the fit by several orders of magnitude, even if the percentage of missing data is small.

We proposed a method that we dubbed ``KARMA'', which provides a general way to perform precise linear regressions with large and incomplete data sets affected by unknown colored noise, and that we applied to mock MICROSCOPE data. The estimation variance is decreased down to the same order of magnitude as the least squares estimator with full data, altered by the natural loss of signal due to the $1/\sqrt{N}$ dependence. The method tends to approach the minimum variance estimator of the available data, by approximating the noise autocovariance with a high order AR model.

Our method uses a weighting of the data relying on the estimation of the shape of the PSD. As a result, the performance of the regression mainly depends on the ability of the autoregressive PSD estimate to recover the part of the spectrum that is responsible for the leakage, which is the high frequency part increasing as $f^{4}$ in the MICROSCOPE case. Although this is not shown here, the method has also been successfully tested in a case where the leaking power comes from a thermal $1/f^{2}$ noise projected onto high frequencies. The AR PSD then accurately fits the low frequency slope and allows us to improve the possible regression of high frequency components. 

In addition, the outputs allow us to evaluate the variance of the estimator from a single estimation. We recover the magnitude of the true precision, equal to $10^{-15}$ in our MICROSCOPE illustration. The variance is not estimated with a better accuracy because of the low frequency bias of the AR PSD estimator.  
This bias depends on the missing data pattern, and more particularly on the length of the longest uninterrupted data segment, as well as the number of long segments. This determines the AR order to be chosen, resulting in a trade-off between the bias and the variance of the PSD estimate.

Concerning the scientific objective of the MICROSCOPE mission, the above discussion demonstrates that based on the current noise model of the accelerometers, we will be able to get a 99\% (resp. 68\%)-confidence level detection of a $3 \times 10^{-15}$ (resp. $1 \times 10^{-15}$) EP violation signal, even in the presence of missing data, for a 20 orbit-measurement session (completed in 1.4 days). The mission should include more than 70 sessions of this type, allowing for a detection at the 99\% level even for an amplitude of $1 \times 10^{-15}$.
This has been done for short and very frequent gaps to represent acceleration peaks or saturations due to MLI or tank crackles, as well as for longer and fewer gaps to simulate telemetry interruptions.

Further developments will concentrate on how to increase the accuracy of the noise PSD estimate, for example by using the AR model to perform missing data imputation. Indeed, although this is computationally more expensive, the AR model can be exploited in a Gaussian process regression approach \cite{Rasmussen06gaussianprocesses} to estimate the missing values. 

Finally, there are two potential limitations to the presented method that can be addressed in further extensions. On the one hand, although the AR model can be a good approximation to any PSD and can be fitted very efficiently, it is still a parametric an thus restrictive model. On the other hand, noise and signal parameters are estimated iteratively but separately, so that each step is done conditionally to the previous one. This may result in a loss of accuracy. As a result, a possible generalization is to use the proposed method as an efficient initialization procedure for a more general regression algorithm that would maximize the full likelihood without any prior noise model.

\appendix
\section{\label{annex:random_case}PSD deformation in the case of random missing data patterns}

We derive here the PSD of the masked data in the case where the gaps positions in the time series are drawn from a uniform distribution.
Let $N$ be the length of the time series, $N_{g}$ the number of gaps, and $n_{b,i}$ the indices indicating the location of the beginning of each gap (such that $w_{n_{b,i}}=0$). Each gap ends at the location $n_{b,i} + dn_{i}$ (we adopt the convention $w_{n_{b,i} + dn_{i}} = 1$). By uniformly distributed, we mean that $n_{b}$ is a random variable following a discrete uniform distribution on the interval $\llbracket 0 , N-1\rrbracket$. We also allow the gap duration $dn$ to be randomly distributed. The window vector is then generated by drawing $N_{g}$ realizations of $n_{b}$ and $dn$.

The probability $P$ to observe a data at a time $n$ is calculated as follows:
\begin{eqnarray}
P & = & \mathrm{P}( w_{n} = 1 ) \nonumber \\ 
& = & \prod_{i=0}^{N_{g}-1} \mathrm{P}( n < n_{b,i} \text{ or } n \geq n_{b,i} + dn_{i} )  \nonumber \\
& = & \left[ 1 - \mathrm{P}( n_{b} \leq n ) +  \mathrm{P}( n_{b} + dn \leq n) \right]^{N_{g}}.
\label{eq:observation_probability}
\end{eqnarray}
The cumulative probability function of $n_{b}$ is given by:
\begin{equation}
\mathrm{P}( n_{b} \leq n ) = \frac{n+1}{N}.
\label{eq:uniform_CDF}
\end{equation}
In the case where the duration of the gaps is fixed (\textit{i.e.} $dn_{i} = dn_{0} \text{ } \forall i$), Eq. (\ref{eq:observation_probability}) gives:
\begin{eqnarray}
\mathrm{P}( w_{n} = 1 ) & = & \left[ 1 - \frac{n+1}{N} + \frac{n-dn_{0}+1}{N} \right]^{N_{g}}  \nonumber \\
& = & \left[ \frac{N-dn_{0}}{N} \right]^{N_{g}}.
\label{eq:observation_probability2}
\end{eqnarray}
Therefore the probability law of $w_{n}$ is a Bernoulli's law of parameter $P$. Its expectation is $\mu_{w} = P$ and its variance is $\sigma_{w}^{2} = P(1-P)$.
We notice that $P$ is independent of time, and the autocovariance function of $\bm{w}$ is simply $R_{w}(n) = \sigma_{w}^{2} \delta(n)$ where $\delta(n)$ is the delta Dirac function.
Then we use Eq. (\ref{eq:observed_PSD}) to calculate the PSD of the masked data:
\begin{equation}
S_{y}(f)   =  P^{2} \cdot S_{z}(f) + P(1-P) \int_{-\frac{f_{s}}{2}}^{\frac{f_{s}}{2}} S_{z}(f') df'.
\label{eq:observed_PSD_MICROSCOPE}
\end{equation}

\section{\label{annex:OLS_variance}Derivation of a simplified equation for the OLS variance in the harmonic case}

We derive here the approximate expression of the variance of the ordinary least squares estimator used in Section \ref{section:impact}. 

We start from Eq. (\ref{equ:OLS_covariance}). In the case of a simple harmonic model, the matrix $A_{w}$ is a column matrix and the covariance formula can be written as:
\begin{equation*}
\mathrm{Var}(\hat{\delta}) = \frac{A_{w}^{\dag} \Sigma_{w} A_{w}}{\left( A_{w}^{\dag}A_{w} \right)^{2}}.
\label{eq:covariance_simple}
\end{equation*}
As reminded in Eq. (\ref{eq:PSD_covariance}), the covariance matrix is diagonalizable in the Fourier space. We keep the same notations in the following.
In addition, we use the fact that the Discrete Fourier Transform (DFT) operator is a Vandermonde matrix (since $M^{\dag}M = N I$ with $I$ the identity matrix), therefore the variance can be rewritten in terms of the DFT of the windowed model $A_{w}$, noted $\tilde{A}_{w} = M A_{w}$:
\begin{equation*}
\mathrm{Var}(\hat{\delta}) = N f_{s} \frac{\tilde{A}_{w}^{\dag} D \tilde{A}_{w}}{\left( \tilde{A}_{w}^{\dag}\tilde{A}_{w} \right)^{2}}.
\label{eq:covariance_simple}
\end{equation*}
By developing this expression we get:
\begin{equation*}
\mathrm{Var}(\hat{\delta}) = \frac{ \sum_{k=0}^{N-1} |{\tilde{A_{w}}}_{k}|^{2} N f_{s}  \hat{S}_{y_{k}}  }{\left( \sum_{k=0}^{N-1} |\tilde{A_{w}}_{k}|^{2} \right)^{2}}.
\label{eq:covariance_simple_developed}
\end{equation*}

For the windowed harmonic model ${A_{w}}_{n} =  w_{n}  \gamma_{\rm{EP}} \sin( 2 \pi n f_{\rm{EP}} / f_{s} + \phi_{\rm{EP}})$, $\tilde{A}_{w}$ is the convolution of the DFT of the window and the DFT of the EP signal. 
In the case of a random window, $|\tilde{A}_{w}|$ usually peaks at the EP frequency with a value of $\gamma_{\rm{EP}} N_{o}/2$ where $N_{o}$ is the number of observed data (where $w_{n}=1$). To simplify the calculations, we neglect the terms at all other frequencies. This amounts to ignoring the leakage of the harmonic signal (but note that the leakage of the noise component is present in $S_{y}$ through equation \ref{eq:observed_PSD}). Furthermore, if we assume that the integration period is an integer multiple of the EP period (\textit{i.e.} there exist an integer $k_{\rm{EP}}$ such that $f_{\rm{EP}} = k_{\rm{EP}} f_{s}/N$), then we have:
\begin{equation*}
   \mathrm{Var}(\hat{\delta}) \approx \frac{ \gamma_{\rm{EP}}^{2} \frac{N_{o}^{2}}{4}  N f_{s} \left( S_{y}(f_{\rm{EP}}) + S_{y}(-f_{\rm{EP}}) \right)  }{\left( 2 \times \gamma_{\rm{EP}}^{2} \frac{N_{o}^{2}}{4}  \right)^{2}},
\label{eq:covariance_simple_approx}
\end{equation*}
where we have made the approximation, valid for large $N$, that the DFT of the autocovariance function in Eq. (\ref{eq:discrete_spectrum}) is equal to the real PSD.
By simplifying we get equation \ref{eq:masked_variance}:
\begin{equation*}
\mathrm{Var}(\hat{\delta}) \approx \frac{2f_{s} N S_{y}(f_{\rm{EP}}) }{N_{o}^{2} \gamma_{\rm{EP}}^{2}}.
\label{eq:masked_variance2}
\end{equation*}
Note that in the case of a complete data set ($w_{n}=1 \text{ } \forall n$) we have $N_{o} = N$ and this formula is more accurate because the model $|\tilde{A}_{w}|$ exactly peaks at $\gamma_{\rm{EP}} N/2$.

\section{\label{annex:state_space} State space equation of an AR model}

We detail here the Kalman equations presented in Section \ref{subsection:weights}.  

The observation matrix, the model matrix and the model noise matrix are defined respectively by:
\begin{eqnarray*}
	H &\equiv & \left( 1,0,\ldots,0 \right)^{T} \\
	F & \equiv & 
\begin{pmatrix}
	0 & 1 & 0 & \hdots & 0  \\
	0 & 0 & 1 & \hdots & 0  \\
	\vdots & \vdots & \vdots & \vdots & \vdots  \\
	0 & 0 & 0 & \hdots & 1  \\
	-a_{p} & -a_{p-1} & -a_{p-2} & \hdots & -a_{1}
\end{pmatrix} \\
G & \equiv & \left( 1 , g_{1}, \ldots , g_{p-1} \right)^{T}.
\end{eqnarray*}
The vector $G$ whose elements are defined by $\epsilon_{n} g_{j} \equiv z_{n+j-1|n} - z_{n+j-1|n-1}$ can be calculated from the AR parameters (see \citet{jones1980maximum}).

The Kalman filter equations in presence of missing data are briefly reviewed here:\\

\textbf{Prediction equation}
\begin{eqnarray*}
\bm{x}(n | n-1) & = & F\bm{x}(n-1 | n-1), \\
\Sigma(n | n-1) & = & F\Sigma(n-1 | n-1)F^{T} + Q,
\label{eq:prediction}
\end{eqnarray*}
where $Q \equiv GG^{T}$. \\

\textbf{Update equation}

The update equation adapted to the missing data case can be formulated as follows:
\begin{eqnarray*}
\Sigma(n | n) & = & w_{n} \left\{ \Sigma(n | n-1) - K(n)H^{T} \Sigma(n | n-1) \right\} \\
              &   & + (1-w_{n}) \left\{ \Sigma(n | n-1) \right\}, \\
\bm{x}(n | n) & = & w_{n} \left\{ \bm{x}(n | n-1) + K(n) \left( z(n) - H^{T}\bm{x}(n | n-1) \right) \right\} \\
              &   & + (1-w_{n}) \left\{ \bm{x}(n | n-1) \right\},
\label{eq:update_observed}
\end{eqnarray*}
where we defined 
\begin{eqnarray*}
K(n) & \equiv & \Sigma(n | n-1) H \left( H^{T} \Sigma(n | n-1) H \right)^{-1}.
\label{eq:K}
\end{eqnarray*}
Note that if the data is not observed at time $n$, the state variance and the state vector are not updated and set equal to the predicted values at previous time.

\begin{acknowledgments}
The authors would like to thank all the members of the MICROSCOPE
Performance team, as well as Sandrine Pires and Jean-Philippe Ovarlez, for fruitful discussions.
This activity has been funded by ONERA and CNES. We also acknowledge the
financial contribution of the UnivEarthS Labex program at Sorbonne Paris
Cit\'e (ANR-10-LABX-0023 and ANR-11-IDEX-0005-02).
\end{acknowledgments}

\bibliography{references}

\end{document}